\documentclass[a4paper,11pt]{article}
\pdfoutput=1 % if your are submitting a pdflatex (i.e. if you have
\usepackage{jinstpub} % for details on the use of the package, please
\title{\boldmath The Mu2e undoped CsI crystal calorimeter}

\author[a]{N.~Atanov,}
\author[a]{V.~Baranov,}
\author[a]{J.~Budagov,}
\author[e]{F.~Cervelli,}
\author[b]{F.~Colao,}
\author[b]{M.~Cordelli,}
\author[b]{G.~Corradi,}
\author[a]{Yu.I.~Davydov,}
\author[e,1]{S.~Di Falco,\note{Corresponding author.}}
\author[b,j]{E.~Diociaiuti,}
\author[e,g]{S.~Donati,}
\author[b,k]{R.~Donghia,}
\author[c]{B.~Echenard,}
\author[b]{S.~Giovannella,}
\author[a]{V.~Glagolev,}
\author[i]{F.~Grancagnolo,}
\author[b]{F.~Happacher,}
\author[c]{D.G.~Hitlin,}
\author[b,d]{M.~Martini,}
\author[b]{S.~Miscetti,}
\author[c]{T.~Miyashita,}
\author[e,f]{L.~Morescalchi,}
\author[h]{P.~Murat,}
\author[e]{E.~Pedreschi,}
\author[e]{G.~Pezzullo,}
\author[c]{F.~Porter,}
\author[e]{F.~Raffaelli,}
\author[b,d]{M.~Ricci,}
\author[b]{A.~Saputi,}
\author[b]{I.~Sarra,}
\author[e]{F.~Spinella,}
\author[i]{G.~Tassielli,}
\author[a]{V.~Tereshchenko,}
\author[a]{Z.~Usubov,}
\author[c]{R.Y.~Zhu}

\affiliation[a]{Joint Institute for Nuclear Research, Dubna, Russia}
\affiliation[b]{Laboratori Nazionali di Frascati dell'INFN, Frascati, Italy}
\affiliation[c]{California Institute of Technology, Pasadena, United States}
\affiliation[d]{Universit\`a ``Guglielmo Marconi'', Roma, Italy}
\affiliation[e]{INFN Sezione di Pisa, Pisa, Italy}
\affiliation[f]{Dipartimento di Fisica dell'Universit\`a di Siena, Siena, Italy}
\affiliation[g]{Dipartimento di Fisica dell'Universit\`a di Pisa, Pisa,
 Italy}
\affiliation[h]{Fermi National Laboratory, Batavia, Illinois, USA}
\affiliation[i]{INFN Sezione di Lecce, Lecce, Italy}
\affiliation[j]{Dipartimento di Fisica dell'Universit\`a di Roma Tor Vergata, Rome, Italy}
\affiliation[k]{Dipartimento di Fisica dell'Universit\`a degli Studi Roma Tre, Rome, Italy}

\emailAdd{stefano.difalco@pi.infn.it}

\abstract{
The Mu2e experiment at Fermilab will search for Charged Lepton Flavor Violating conversion of a muon to an electron in an atomic field. The Mu2e detector is composed of a tracker, an electromagnetic calorimeter and an external system, surrounding the solenoid, to veto cosmic rays. 
The calorimeter plays an important role to provide: a) excellent particle identification capabilities; b) a fast trigger filter; c) an easier tracker track reconstruction. 
Two disks, located downstream of the tracker, contain 674 pure CsI crystals each. Each crystal is read out by two arrays of UV-extended SiPMs.  The choice of the crystals and SiPMs has been finalized after a thorough test campaign. 
A first small scale prototype consisting of 51 crystals and 102 SiPM arrays has been exposed to an electron beam at the BTF (Beam Test Facility) in Frascati. Although the readout electronics were not final, results show that the current design is able to meet the timing and energy resolution required by the Mu2e experiment.
}


\arxivnumber{1801.02237} % only if you have one

\collaboration[c]{on behalf of Mu2e collaboration}

\proceeding{Calorimetry for the high energy frontier
(CHEF17)\\ 2 - 6 October 2017\\   Lyon, France}

\begin{document}
\maketitle
\flushbottom

\section{Introduction}
\label{sec:intro}
The goal of the Mu2e experiment~\cite{TDR} at Fermilab is to look for neutrinoless conversion of a muon into an electron in the field of an aluminum atom. This two-body process produces a monoenergetic conversion electron of 104.967 MeV. The experimental apparatus has been designed to reduce the number of background events mimicking the conversion electron to $\sim$0.5 in 3 years of data taking. Using $\sim10^{18}$ muons stopped in the aluminum target, Mu2e aims to improve by a factor $10^4$ the current experimental sensitivity ~\cite{sindrum2} on muon conversion.

\section{The Mu2e experiment}
\label{sec:mu2e}

\begin{figure}[htbp]
\centering
\includegraphics[height=5cm]{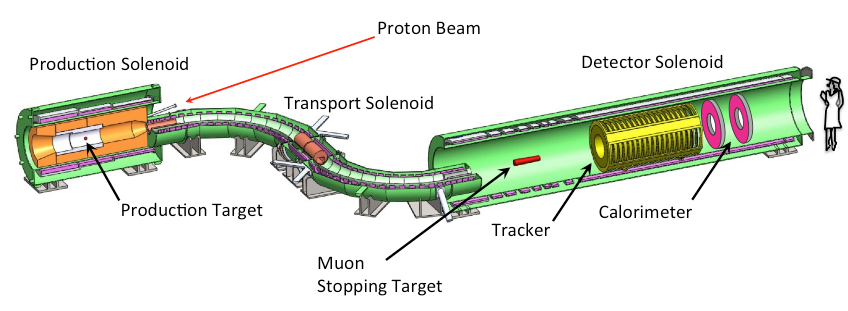}
\caption{\label{fig:mu2e} The Mu2e experiment.}
\end{figure}

In the Mu2e experiment (figure~\ref{fig:mu2e}) an 8 GeV proton beam enters into the production solenoid where it interacts with a tungsten target. Charged particles resulting from the interaction (mostly pions and kaons together with their decay products) are guided by a graded magnetic field towards a transport solenoid that selects negative particles (muons) with momentum $\sim$50 MeV which are transferred to the detector solenoid. Here, muons are absorbed by the aluminum target and the products of muonic aluminum decays (lifetime 864 ns) are directed by a graded field to a constant 1 T field region where the tracker and the calorimeter are located. 
The bunched structure of the proton beam helps to suppress prompt backgrounds due to radiative pion captures, muon decays in flight and beam electrons, by looking only for electrons arriving 700 ns after each bunch.
 
The Mu2e tracker consists of about 21000 straw tubes grouped into 18 measurement stations distributed over a distance of $\sim$3 m. Each straw tube is
read on both sides by TDCs and ADCs.
The core of the momentum resolution for 105 MeV electrons is expected to be better than 180 keV/c.
Cosmic muons backgrounds ($\delta$ rays, muon decays or misidentified muons) are suppressed by a Cosmic Ray Veto system of plastic scintillating counters that covers the whole detector solenoid and half of the transport solenoid. 
Since a veto efficiency of 99.99\% (corresponding to $\sim$ 1 background event in 3 years) is not sufficient to fulfill the experimental requirements, a particle identification, obtained from the crossed information of tracker and calorimeter information, has to provide an additional rejection factor of 200.
Particle identification cannot reject electrons produced by cosmic muons, leaving an irreducible background of 0.2 expected events in 3 years.

%%Due to electrons produced by cosmic muons, a final irreducible background of 0.2 events in 3 years is expected.

\section{The Mu2e electromagnetic calorimeter}
\label{sec:ecal}

The Mu2e electromagnetic calorimeter (ECAL) is expected to:
a) identify, together with the tracker, the conversion electrons with an efficiency higher than 90\% and to reject the mimicking cosmic muon background by a factor higher than 200;
b) provide a trigger in less than $\sim$3 ms with an efficiency on conversion electron events higher than 90\% and with a background rejection power higher than 100;
c) help the tracker pattern recognition to improve track reconstruction efficiency and robustness against background events.

Simulation \cite{TDR} shows that these requirements are fulfilled if the calorimeter is able to reconstruct conversion electrons with an energy resolution better than 10\%, a time resolution better than 500 ns and a position resolution better than $\sim$ 1 cm.

\begin{figure}[htbp]
\centering
\includegraphics[height=5.cm]{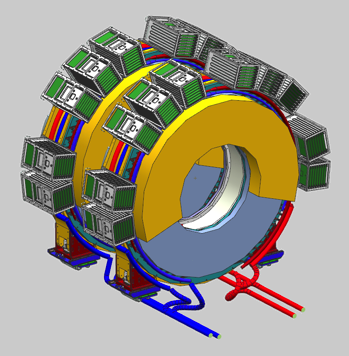}
\qquad
\includegraphics[height=3cm]{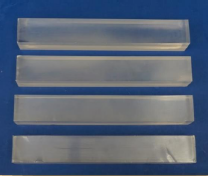}
\qquad
\includegraphics[height=3cm]{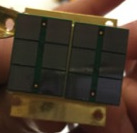}
\caption{\label{fig:ecal} The Mu2e electromagnetic calorimeter (left), pure CsI crystals (center) and the two SiPM arrays reading a crystal (right).}
\end{figure}

Fig. \ref{fig:ecal} shows a picture of the Mu2e calorimeter: two annular disks of undoped CsI crystals are placed at a relative distance of $\sim 70$ cm, half of the average pitch of the helix of a conversion electron inside the magnetic field. 
The disks have an inner radius of 37.4 cm and an outer radius of 66 cm.
Each disk contains 674 undoped CsI crystals of $20\times3.4\times3.4$ cm$^3$. The granularity requirement results from optimization of light collection for photosensors, from minimization of particles pileup and from the needed time and energy resolution. 
Each crystal is wrapped with 150 $\mu$m of Tyvek 4173D and read out by two arrays of UV-extended silicon photomultiplier sensors (SiPM). The SiPM signal is amplified and shaped in the Front-End Electronics (FEE) board located on its back. Digital electronics sampling the signals at 200 MHz is located in crates distributed around the disks.

Quality tests on a set of pure CsI crystals from Amcrys, Saint Gobain and SICCAS  (fig.\ref{fig:ecal}) have been performed at Caltech and at the INFN Laboratori Nazionali di Frascati (LNF).
These tests concerned light yield ($>$100 p.e./MeV when measured with a 2'' UV extended EMI PMT), longitudinal response uniformity (better than 5\%), decay time $(\sim 30$ ns), slow component contamination due to crystal impurities (less than 25\%), light output reduction due to a total ionizing dose of 100 krad (lower than $40\%$), light output and longitudinal response uniformity reduction due to a total fluence of $9\times 10^{11}$ n/cm$^2$ (negligible) and radiation induced readout noise at 1.8 rad/h (equivalent to less than 600 KeV).
The last two vendors were found to have a lower slow component contamination.

SiPM sensor arrays (fig.\ref{fig:ecal}) are obtained by assembling two series of 3 monolithic sensors. The two series are connected in parallel by the Front End electronics to have a x2 redundancy.
The series connection reduces the global capacitance, reducing the signal decay time to less than 100 ns and minimizes output current and power consumption. 
Each monolithic sensor has an active surface of 6x6 mm$^2$ and is UV-extended with a photon detection efficiency (PDE) at the CsI emission peak ($\sim$315 nm) of $\sim$ 30\%.
Quality tests on a set of SiPM arrays from Advansid, Hamamatsu and SENSL have been performed at INFN Pisa and Frascati. 
The tests concerned gain at the operating voltage (3V above the breakdown voltage: $>10^6$ integrating charge on a 150 ns gate), Photon Detection Efficiency ($>$ 20\%), uniformity of the sensor array (breakdown voltage spread <0.5\%, dark current spread <15\%), dark current at 0$^\circ$ C after an irradiation with $3\times 10^{11}$ neutrons/cm$^2$ 1 MeV equivalent ($<2$ mA) and mean time to failure ($>6\cdot10^5$ h).
The Hamamatsu sensors were found to have a better timing performance. Their light yield, when coupled in air with a CsI crystal, is $\sim$ 20 p.e./MeV and
the noise contribution to energy resolution is $\sim$ 100 keV.

\section{Beam test of a small scale prototype}

\begin{figure}[htbp]
\centering
\includegraphics[height=4cm]{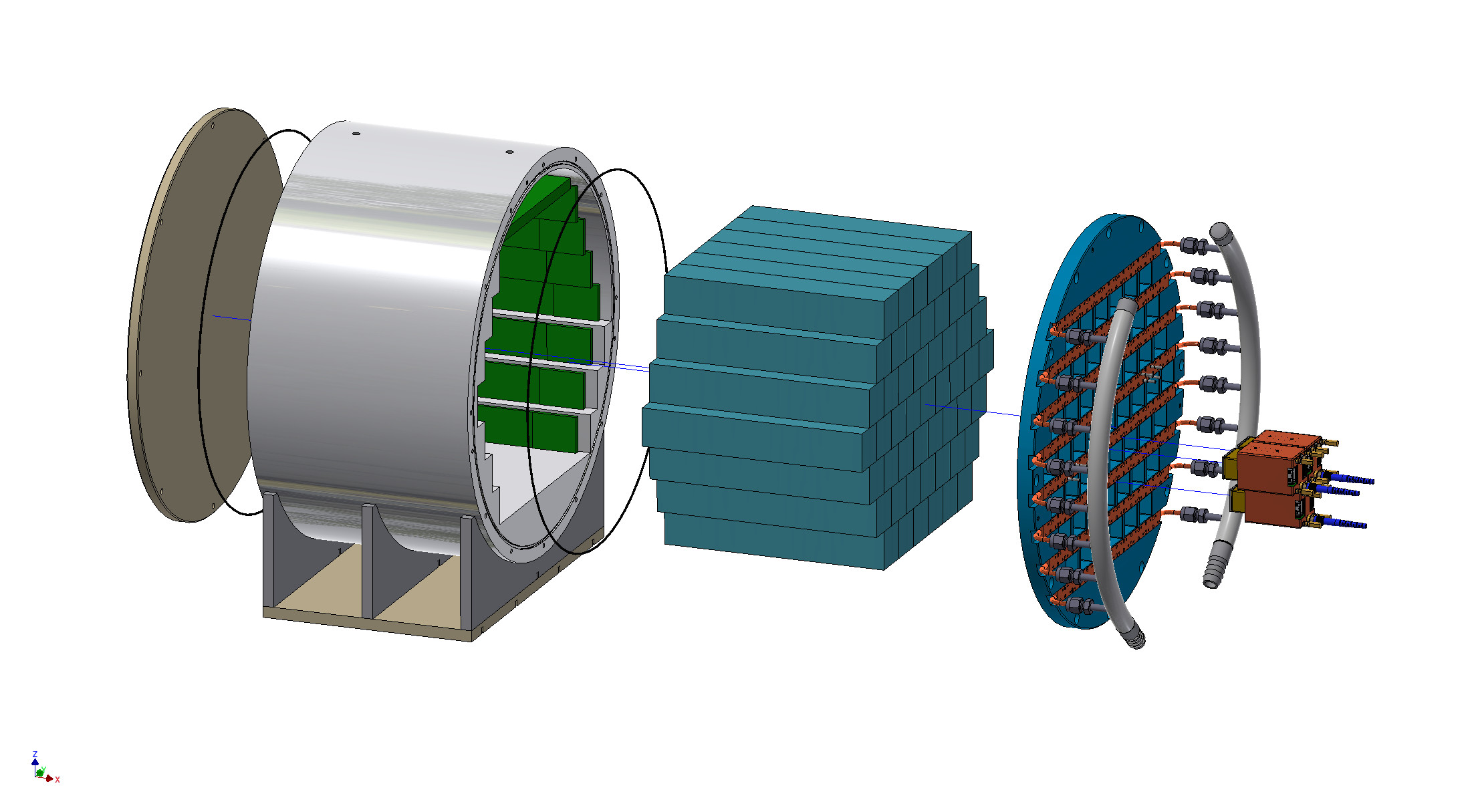}
\qquad
\includegraphics[height=3cm]{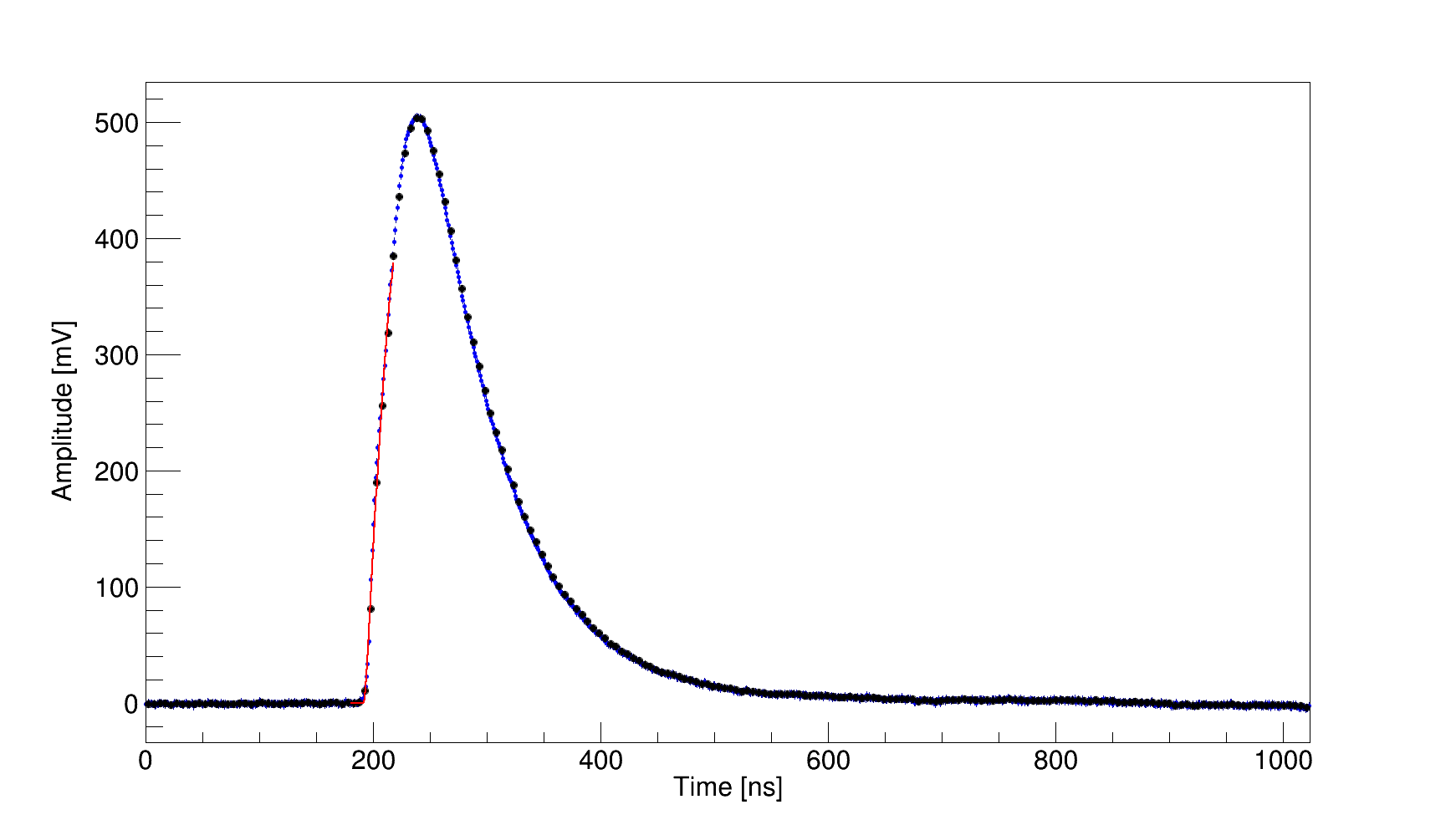}
\caption{\label{fig:module0} The calorimeter Module 0 (left) and example of a digitized waveform acquired at the beam test(right).}
\end{figure}

A calorimeter prototype (``Module 0'') has been built
using 51 crystals and 102 SiPMs, which previously passed the quality tests. The module has been tested in May 2017 at the Beam Test Facility in Frascati.
Fig. \ref{fig:module0} shows Module 0 and its components: on the back of the SiPMs are located the SiPM copper holders, the shaping amplifier boards and the copper cooling pipes used to keep the SiPM temperature at 20$^\circ$ C\footnote{This was done to partially validate also the cooling system design. A final test at 0$^\circ$ C in vacuum will be performed in the future.}. 

Final calorimeter electronics were not yet available at beam test time. Front end amplifiers were shaping the signal (fig. \ref{fig:module0}) with the final rise time ($\sim$ 20 ns) but with a larger total width  (400 ns instead of 150 ns). Two commercial 1 GHz digitizers from CAEN have been used to sample via ADCs a total of 64 channels: 1 SiPM for each of the 51 crystals, a second SiPMs for the 7 central crystals. The remaining 6 channels were used for trigger signals: 2 thin, narrow plastic scintillators located in front of the calorimeter, the accelerator clock, 2 PMTs reading a  cosmic counter on top of the calorimeter and the laser pulse used to illuminate the central crystal.
The use of preliminary electronics had some undesired drawbacks producing an uncorrelated noise of $\sim 300$ keV/channel and a time jitter between different digitizer chips (controlling 8 channels) of $\sim$ 1 ns.

\begin{figure}[htbp]
\centering
\includegraphics[height=3.7cm]{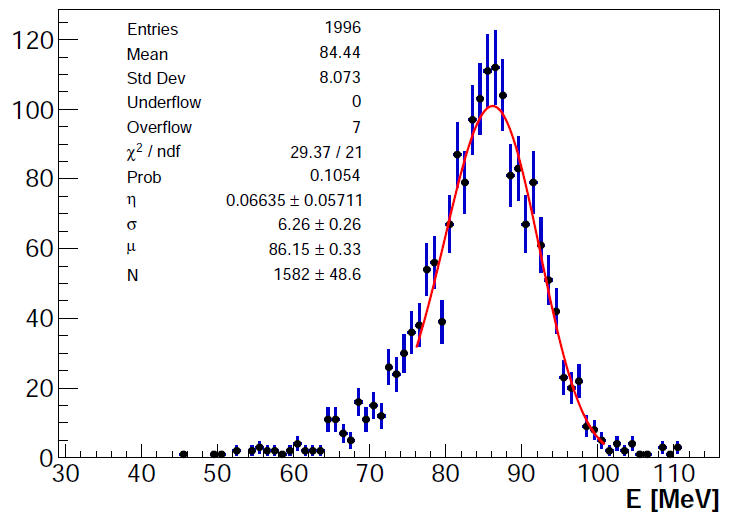}
\qquad
\includegraphics[height=4cm]{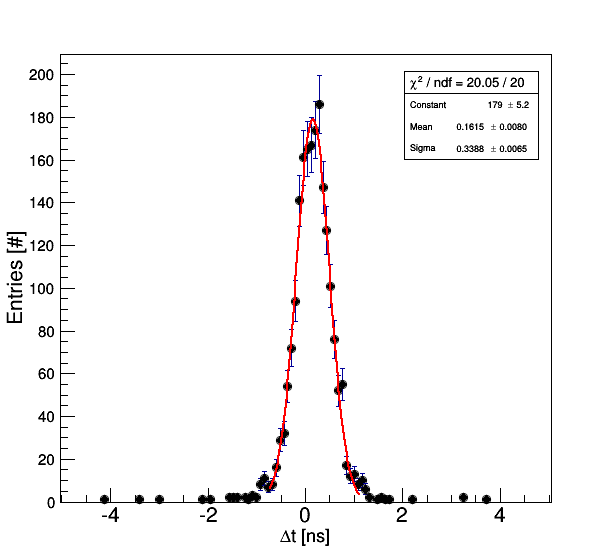}
\caption{\label{fig:res} Energy resolution (left) and time resolution with 200 MHz sampling frequency (right) for 100 MeV electrons at $50^\circ$ .}
\end{figure}

A 100 MeV electron beam at 0$^\circ$ has been used to equalize the channels response. The same 100 MeV electron beam with an impact angle of 50$^\circ$ in the horizontal plane has been used to evaluate the time and energy performances in the same conditions as the Mu2e experiment\footnote{In Mu2e the 105 MeV electrons produced in the stopping target will impact the calorimeter with an angle ranging from 45$^\circ$ to 60$^\circ$.}.
After pedestal subtraction, the charge deposited in each crystal has been obtained by integrating the signal waveform in a 250 ns gate starting from the waveform leading edge. A simple clustering obtained by adding the information from the hit  crystal and from crystals in the first four rings surrounding the hit crystal has been used to get the total shower charge.  The result is shown in fig.\ref{fig:res}: the bulk of the distribution has a $\sigma\sim$7\%. The tail on the left is due to longitudinal leakage. 
  
After resampling the waveform at 200 MHz, the particle arrival time has been obtained by fitting the leading edge of the distribution (fig. \ref{fig:module0}) with a log-normal function and imposing a threshold depending on the peak amplitude (constant fraction).
The time resolution has been estimated using the difference of times from the sensors reading the crystal with the highest energy deposit. The sigma of the time difference (fig.\ref{fig:res}) is $\sim$340 ps, corresponding to a single channel resolution of 240 ps and a resolution of 170 ps on the average of the two sensors. This is just a preliminary estimate: other effects, such as the synchronization of different digitizer boards or the dependence of the largest energy deposit in a crystal on the impact point and on the 3-dimensional angle, need to be tested in a future beam tests using final electronics.
  
The present beam test shows that energy and time resolution of the Mu2e calorimeter stay well within the required 10\% and 500 ps.

\section{Conclusions and outlook}
The Mu2e calorimeter design is now mature: crystals and SiPMs have been chosen and their production will start in 2018. Calorimeter construction is expected to be completed by the beginning of 2020.
Monte  Carlo simulation, supported by test beam results, shows that
the calorimeter design meets the requirements on muon identification, seeding of track reconstruction and trigger selection needed for the Mu2e experiment. 

\acknowledgments

We are grateful for the vital contributions of the Fermilab staff and the technical staff of the participating institutions.
This work was supported by the US Department of Energy; 
the Italian Istituto Nazionale di Fisica Nucleare;
the Science and Technology Facilities Council, UK;
the Ministry of Education and Science of the Russian Federation;
the US National Science Foundation; 
the Thousand Talents Plan of China;
the Helmholtz Association of Germany;
and the EU Horizon 2020 Research and Innovation Program under the Marie Sklodowska-Curie Grant Agreement No.690385. 
Fermilab is operated by Fermi Research Alliance, LLC under Contract No.\ De-AC02-07CH11359 with the US Department of Energy, Office of Science, Office of High Energy Physics.
The United States Government retains and the publisher, by accepting the article for publication, acknowledges that the United States Government retains a non-exclusive, paid-up, irrevocable, world-wide license to publish or reproduce the published form of this manuscript, or allow others to do so, for United States Government purposes.


\begin{thebibliography}{99}

\bibitem{TDR}
Mu2e Collaboration (L. Bartoszek et al.), \emph{Mu2e Technical Design Report}, arXiv:1501.05241.

\bibitem{sindrum2}
SINDRUM II Collaboration (W.H. Bertl et al.), \emph{A Search for muon to electron conversion in muonic gold}, \emph{Eur.Phys.J.} {\bf C47} (2006) 337-346.




\end{thebibliography}
\end{document}